# Progressive Disclosure: Designing for Effective Transparency

**Aaron Springer**
University of California Santa Cruz
Santa Cruz, CA, USA
alspring@ucsc.edu

**Steve Whittaker**
University of California Santa Cruz
Santa Cruz, CA, USA
swhittak@ucsc.edu

**ABSTRACT**
As we increasingly delegate important decisions to intelligent systems, it is essential that users understand how algorithmic decisions are made. Prior work has often taken a technocentric approach to transparency. In contrast, we explore empirical user-centric methods to better understand user reactions to transparent systems. We assess user reactions to global and incremental feedback in two studies. In Study 1, users anticipated that the more transparent incremental system would perform better, but retracted this evaluation after experience with the system. Qualitative data suggest this may arise because incremental feedback is distracting and undermines simple heuristics users form about system operation. Study 2 explored these effects in depth, suggesting that users may benefit from initially simplified feedback that hides potential system errors and assists users in building working heuristics about system operation. We use these findings to motivate new progressive disclosure principles for transparency in intelligent systems.

**ACM Classification Keywords**
H.5.m. Information interfaces and presentation (e.g., HCI):

**INTRODUCTION**
Machine learning algorithms underlie the many intelligent systems we routinely use. These systems provide information ranging from routes to work to recommendations about criminal parole [3,8]. As humans with limited time and attention, we increasingly defer responsibility to these systems with little reflection or oversight. Nevertheless, intelligent systems face mounting criticisms about how they make decisions; criticisms that are exacerbated by recent machine learning advances like deep learning that are difficult to explain in human-comprehensible terms. Major public concerns have arisen following demonstrations of bias in algorithmic systems with regards to gender, race, and other characteristics [9,73,76]. These issues have led to calls for transparency as a solution to the unintelligible algorithms that impair the adoption of intelligent systems [31,61,80].

Algorithmic transparency is needed for many reasons. Greater transparency potentially increases end user control and improves acceptance of complex algorithmic systems [41]. It can also promote user learning and insight from complex data, as humans increasingly work with complex inferential systems for analytic purposes [41,69]. Transparency can also enable oversight by system designers. Without such transparency it may be unclear whether an algorithm is optimizing the intended behavior [32,49], or whether an algorithm accidentally promotes negative, unintended consequences (e.g. filter bubbles in social media; [10,62]). Given these current issues, it is increasingly possible that transparency, i.e. "a right to explanation", may become a legal requirement in some contexts [28]. These issues have led researchers to argue that machine learning must be 'interpretable by design' [1] and that transparency is essential for the adoption of intelligent systems, e.g. for medical diagnoses [31,80].

While such calls for transparency are well-motivated, it remains unclear exactly how to enact them in practice. Extensive research about operationalizing transparency has emerged in the machine learning community but no clear consensus has resulted [1,21,79]. Deciding exactly how to implement transparency is difficult—there are numerous implementation trade-offs involving accuracy and fidelity. Making a complex algorithm understandable to end users might require simplification, which often comes at the cost of reduced accuracy of explanation [44,69]. For example, methods have been proposed to explain neural network algorithms in terms of more traditional machine learning approaches, but these explanations necessarily present approximations of the actual algorithms deployed [51]. These studies often approach transparency from a technical perspective: "what is possible from an algorithmic standpoint?" rather than "what does the user need?"

However, some recent empirical studies attempt to examine the effects of transparency on users. But these studies reveal puzzling and sometimes contradictory effects. In some settings there are expected benefits: transparency improves algorithmic perceptions because users better understand system behavior [40,41,47]. But in other circumstances, transparency has other quite paradoxical effects. Transparency may erode confidence in a system, with users trusting it less because transparency led them to question the system even when it was correct [47]. Providing system explanations may also undermine user perceptions when users lack the attentional capacity to process complex explanations, for example while they are executing a demanding task [12,81]. Overall, these results indicate mixed evidence for the benefits of transparent systems.

The above research suggests that we have yet to identify the appropriate interaction paradigms to present transparency. Machine learning research communities are forging ahead with foundational research on how to generate transparent systems [79] but studies often stop short of actually testing these systems with users [46,48]. Evaluation is critical because, as we have seen, user reactions to transparency show quite contradictory results [12,41,47]. Some research suggests that the way we present transparency may account for these contradictory results [26,40]. Our research seeks to bridge this gap between generating explanations and user-centric presentation. In two studies, we explore users' direct reactions to a transparent personal informatics system that interprets their emotions. We examine user preferences for different forms of transparency by comparing two versions of a working system: one that provides detailed incremental feedback about the underlying algorithm and another system version that provides holistic global feedback. We also examine the role of cognitive load in explaining these preferences, and explore problems with current presentations of transparency. We address the following research questions:

- RQ1: Do users prefer to use more transparent systems? Do they prefer systems providing detailed incremental feedback or those that offer global transparency? (Study 1)
- RQ2: Does cognitive load and distraction play a role in preferences for transparency? (Study 1)
- RQ3: What problems must be mitigated to support effective transparency? (Studies 1 and 2)

CONTRIBUTION: Much recent work on transparency has focused on technical explorations of self-explanatory systems. In contrast, here we take an empirical user-centric approach to better understand how to design transparent systems. Two studies provide novel data concerning user reactions to systems offering incremental vs global transparency information. In Study 1 users anticipated that an incremental system would perform better, but retracted this evaluation after experience with the system. Qualitative data suggest this may be because incremental feedback can be distracting and potentially undermines simple heuristics users form of system operation. Study 2 explored these effects in more detail suggesting that users may benefit from simplified feedback that hides potential system errors and assists users in building working heuristics about system operation. We use these data to motivate new progressive disclosure principles for presenting transparency in intelligent systems.

## RELATED WORK

### Folk Theories of Algorithms

A wealth of prior work has explored issues surrounding algorithm transparency in the commercial deployments of systems for social media and news curation. Social media feeds are often curated by algorithms that may be invisible to users (e.g., Facebook. Twitter, LinkedIn). At one point, most users were unaware that Facebook newsfeeds were not simply all the posts that their friends made [24]. These users reacted in surprise and sometimes anger when they were shown the posts that were missing from their newsfeed. Later research shows that many users of Facebook develop 'folk theories' of their social feed [22], which are imprecise heuristics about how the system works, even going so far as to make concrete plans based upon their folk theories. This work also showed that making the design more transparent or seamful, allowed users to generate multiple folk theories and more readily compare and contrast between them [22].

Other work has illustrated issues regarding incorrect folk theories in the domain of intelligent personal informatics systems, showing specific challenges in how users understand these systems. Users are prone to blindly believing outputs from algorithmic systems, a phenomena referred to as algorithmic omniscience [23,36,74] and automation bias [17,56]. For example, KnowMe [78] is a program that infers personality traits from a user's posts on social media based on Big Five personality theory. KnowMe users were quick to defer to algorithmic judgment about their own personalities, stating that the algorithm is likely to have greater credibility than their own personal statements (e.g., "...At the end of the day, that's who the system says I am..."). Similar results were shown in [36], showing that participants expected intelligent personal informatics systems to serve as ground truth for their experiences and even attributed superhuman qualities to these devices, e.g., "...[it] could tell me about an emotion I don't know that I am feeling...". Other experiments indicate the risk of such trust, showing that users may believe even entirely random system outputs as moderately accurate [74]. Similarly, giving users placebo controls over an algorithmic interface shows corroborating results [77]; users with placebo controls felt more satisfied with their newsfeed. Without a standard of transparency in intelligent systems, it may be easy to deceive end-users into believing they are using a real system; this is a dangerous proposition when apps can be so easily distributed.

**Transparency**

There is a long history of studying transparency and intelligibility in automated systems [6]. However, the results have mixed and often indicate contradictory effects on user perceptions. Many experiments have indicated that transparency improves user perceptions of the system [20,46]. Others have shown that interventions that simply show prediction confidence improve users system perceptions [4]. In extreme cases, animations that simulate transparency can cause users to be overconfident about systems even when they err [26].

Other studies show less positive effects for user perceptions of a system. Participants who completed an experiment using a hypothetical transparent system were led to question the system an increased amount, resulting in worse agreement with the system [47]. However, the effect may be opposite for high certainty systems—transparency may only result in higher user agreement. Muir and Moray conclude that any hint of error in an automated system will decrease trust [57]. More recent work indicates other effects; explanations of how a system is working may lead to increased trust [40] but further explanations may be harmful to user perceptions. However, these effects are dependent upon the amount of expectation violation that a user experiences. User expectation violation follows an event where a system behaves in a way that a user did not expect [40,74,77]. Ideally transparency should build user confidence in a system, whether or not the user is experiencing expectation violation. However, it seems the research communities have not found the correct interaction paradigms to achieve this.

Recently, the machine learning community has begun grappling with issues of transparency and explainability. This seems due to the rise of more inscrutable methods like deep learning as well as legal requirements arising from the European Union's GDPR. Some machine learning models are "inherently understandable" such as linear models and Generalized Additive Models [50,79]. These understandable models can be "explained" to users simply through the linear contributions of their features. Other algorithms such as deep neural nets and random forests are inscrutable, and it is nontrivial to explain how input features match to output predictions [79]. Many attempts have been made to make these inscrutable algorithms understandable. These rely on approximating the inscrutable algorithm through a simple local or linear model than can be explained to the end user [51,69]. However, even with these methods that result in "inherently understandable" models, there is no clear consensus how to convey these models to users in an understandable way. Many such attempts at transparency are not tested with users or simulate user studies [5,55,58]. Without real user feedback, we cannot hope to operationalize transparency in ways that positively impact users.

**Explanation and Persuasion Theory**

People interact with computers and intelligent systems in ways that mirror how they interact with other people [60,68]. Given that transparency is essentially an explanation of why a model made a given prediction, we can turn to fields such as psychology and sociology for guidance about operationalizing explanations. These fields have a long history of studying explanation. Hilton shows that causal explanation takes the form of conversation and thus is governed by the common-sense rules of conversation [34] such as Grice's maxims [29]. In addition, when explanation is needed and a communication breakdown occurs this is remedied by a phenomena known as *conversational repair*. Conversational repair is interactional, participants in the conversation work together to achieve mutual understanding; this often happens in a turn-by-turn structure with repeated questions and clarifications [71]. These theories would indicate that we should operationalize transparency in ways that fit human communication and repair strategies.

Additionally, we see parallels between how people interact with intelligent systems and persuasion theory. The Elaboration Likelihood Model (ELM) is a dual process model of persuasion [64]. The ELM posits that two parallel processes are engaged when a person evaluates an argument, similar to Kahneman's conception of System 1 and 2 thinking [38]. The central processing route involves careful consideration of the argument and complex integration into a person's beliefs. The central route is often engaged in high stakes decisions. The peripheral route in contrast focuses on heuristic cues such as the attractiveness of the speaker, the person's current affect, the number and length of the arguments, and other cues not directly related to the content of the argument. Prior work on intelligent systems seems to align with this dual process model [40], people understand systems through peripheral routes if their expectations are met, only engaging in central processing when their expectations are violated. This is also demonstrated this in the context of Google search suggestions; where users felt the cost of processing explanations outweighed their benefits [12]. Transparency needs to be operationalized in ways that allow users to understand transparency through both cursory heuristic routes and also through focused effort.

**Emotional Analytics**

Our focus is on how users interact with an intelligent personal informatics systems which are being increasingly deployed within commercial [83–85] and research domains [7,25,35,52,52,82]. These systems track how a person operates on some dimension, whether physical, emotional, or mental, and then suggest improvements to this behavior through customized feedback and recommendations [35,66]. Such data potentially allows users to analyze and modify their behaviors to promote well-being [15,16,37].

Emotional analytics is a fruitful domain for transparency research for many reasons. In contrast to other work that presents hypothetical scenarios in which participants read about or watch algorithmic deployments and decisions [26,47], our aim was to have users experience the algorithm in situ, as it directly made decisions about their own data [39]. One important characteristic of emotional interpretation is that users are knowledgeable about the status of their own feelings and experiences, allowing them to directly compare algorithmic interpretations with their own personal evaluations of their emotional experiences. This contrasts with other applications of smart algorithms, such as medical diagnoses. In these complex realms regular users might be less able to interpret the results of algorithmic interpretations. In addition, emotion is highly variable between individuals and previous research demonstrates difficulty in accurately predicting emotion from text [45,67]. This allows us to closely examine how users understand a system under varying degrees of error.

**RESEARCH SYSTEM: E-METER**

We wanted to test users' reactions to a transparent system that actively interpreted their own data. We developed a working system called the E-meter that uses textual entries to predict emotion. The E-meter (Figs 1,2) presents users with a web page showing a system depiction, a short description of the system, instructions, and a text box to write in. The system was described as an "algorithm that assesses the positivity/negativity of [their] writing".

The algorithm underlying emotion detection worked in the following way: each word that was written by the user was checked for its positive/negative emotion association in our model. If it was found in the model, the overall mood rating in the system was updated. This constitutes an incremental linear regression that recalculates each time a word is written.

**Machine Learning Model**

As we outlined above, current processes for explanation of inscrutable models such as deep neural networks involve approximating the inscrutable model by a simpler, often linear, model [79]. Therefore, we focus on a linear model so that our transparency can be operationalized in a way that is faithful to current research. While this model may be less accurate than those involving other methods, it nevertheless gave us presentational control. It allowed us to directly visualize important elements for users that explained the algorithm's underlying operation.

Emotion predictions for users' experiences were generated using a linear regression model trained on text from the EmotiCal project [35,75]. In EmotiCal, users wrote short textual entries about daily experiences and directly evaluated their mood in relation to those experiences. This data gave us a gold-standard supervised training set on which to train our linear regression. We trained the linear regression on 6249 textual entries and mood scores from 164 EmotiCal users. Text features were stemmed using the Porter stemming algorithm [65] and then the top 600 unigrams were selected by F-score, i.e. we selected the 600 words that were most strongly predictive of user emotion ratings. Using a train/test split of 85/15 the linear regression tested at $R^2 = 0.25$; mean absolute error was .95 on the target variable (mood) scale of (-3,3). In order to implement this model on a larger range for the E-meter, we scaled the predictions to (0,100) to create a more continuous and variable experience for users. The mean absolute error of our model indicates that the E-meter will, on average, err by 15.83 points on a (0,100) scale for each user's mood prediction.

*Document-level version*: As users wrote, the E-meter showed the system's global interpretation of the emotion of their writing. If the overall text was interpreted as positive, the meter filled the gauge to the right and turned more green (Fig 2); if the text was interpreted negatively, the gauge was emptied to the left and turned more red (Fig 1). This continuous scale feedback represents the coarse and global information that many machine learning systems currently display. These systems give an overall rating but don't allow the user insight into the detailed workings of the algorithm.

*Word-level version*: In contrast to the document-level version, the word-level condition provided fine-grained transparency. We operationalized such transparency by highlighting the mood association of each word in the model; if a word is highly associated with positive mood then it will be highlighted green, a word associated with negative mood will be highlighted orange or red. The word-level version showed immediate incremental feedback about how the system interprets each word they type. In this version, individual words are highlighted and color coded according to how the underlying algorithm interpreted that word's affect. This incremental feedback allows users to see how each individual word they write contributes to the overall E-meter rating. Furthermore, words remain highlighted as users continue to type allowing them to assess each word's relative contribution to the overall score.

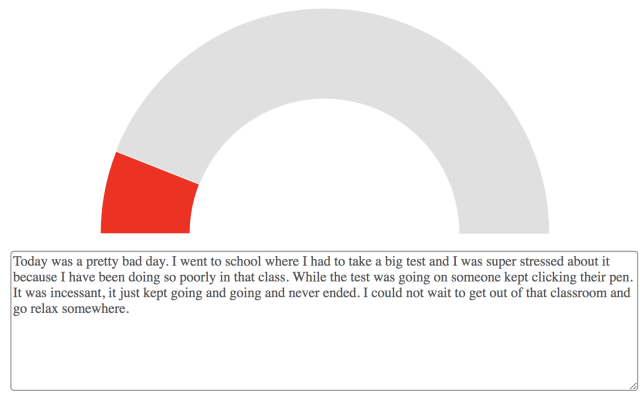

Figure 1: E-meter Document-Level Feedback Condition

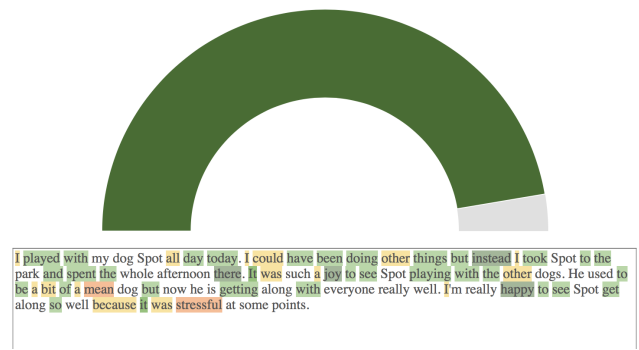

Figure 2: E-meter Word-Level Feedback Condition

This form of transparency offers users insight into the underlying word-based regression model driving the E-meter visualization; it depicts how the regression model correlates each word with positive or negative emotion to arrive at an overall weighting for the entire text that the user has entered. The fact that the visualization is persistent also allows users to reexamine what they have written, reconciling the overall E-meter rating with the fine-grained word-level connotations.

We could have operationalized transparency in other ways. Other researchers have operationalized transparency through natural language explanations [40] and diagrams [47]. However, in our case we can convey key aspects of the underlying system through word highlighting. In addition, our operationalization allows the answering of counterfactual questions, an important part of explanation [54,79]. Highlighting the text helps directly convey to the user what drives the algorithm and gives clear clues about the underlying linear model. In addition, by varying the colors of the highlighting we also show how the model is interpreting the specific words.

## STUDY 1

### Method

Participants experienced both versions of the E-meter system and after using each version they were asked a series of follow-up questions. The study design was counterbalanced; half the users experienced the document-level system version first.

*Users*
We recruited 100 users to test the E-meter system who had previously passed a short mental health screening (PGWBI) [30]. Users were recruited from Amazon Turk and paid $3.33. The evaluation took 14.68 minutes on average. This study was approved by an Institutional Review Board. Following prior methodological recommendations in [18] we eliminated 26 respondents based on their responses to open ended questions, leaving us with a sample of 74 users.

*Measures*
Before actually using the system, participants saw simulations of both versions of the system and were asked to *predict accuracy* for each. The first survey question was preceded by animations simulating both versions of the system as filler Latin text was typed. We used Latin text because we wanted high-level system comparisons from users. If we had shown an animation of typed English, the users could have made their assessments based on reactions to specific words. Following the animations we asked the *predicted accuracy* question "This program evaluates the positivity/negativity of emotional experiences that users write about. How accurate or inaccurate do you think this program would be for you? The program works with English also." Users then provided qualitative explanations for their ratings—"Please give 2 or more reasons for the accuracy ratings you made on the previous page." After this they began the writing activity: Users were presented one of the two system version along with the instructions "Please write at least 100 words about an emotional experience that affected you in the last week." Following system experience, users completed the TLX *workload* assessment [33]. Users then answered the questions: "How positive or negative did you feel your writing was?" (*subjective affect*), "How positive or negative did the E-meter assess your writing to be?" (*system affect*), "How accurate or inaccurate was the E-meter in its assessment of your writing?" (*retrospective accuracy*), "How trustworthy or untrustworthy did you find the E-meter system?" (*subjective trust*).

Users then repeated this process for the other system version. After using both versions, users answered a final *experience-based system preference* question "If you were to use the E-meter again, which system would you prefer?". They then supplied reasons for this: "Please give 2 or more reasons for the choice you made above". Questions and procedure were carefully piloted and had been used before in multiple prior studies.

### Results

Overall, the median user found the E-meter to be 'Slightly Accurate' and 'Slightly Trustworthy', both distributions were bimodal and there was no difference in conditions (p = .24, p=.41).

*Word-level transparency is predicted to be more accurate before usage:* Before any hands-on experience with the system, participants generated *predicted accuracy* judgments for both word-level and document-level transparency. We statistically compared the difference in these system evaluations using a paired t test. Participants anticipated greater accuracy for the word-level system, t(73)=5.452, p=0.022, although the effect was small and means were 4.24 and 4.57 respectively. Qualitative user comments supported these anticipated benefits. On being asked to explain why they had word-level transparency to be more accurate, users drew attention to the benefits of incremental low-level color-coded feedback giving a clearer sense of how the system was operating, and boosting their confidence that the system was operating appropriately. Participant 73 wrote "*The second one had a legend with it and actually changed the color of the words I would have written. ... It was also more catchy and the colors stood out*

*to me."* In the same vein, participant 50 wrote: *"One meter is more transparent than the other. I can see how it works. I feel more confident in knowing exactly how it comes up with its answers. I tend to think it is more reliable."* Overall then, before actually using either transparency version, users anticipated that a system offering word-level transparency would be more accurate.

Recall that after experiencing each system version, we asked users for *retrospective accuracy*, and a final *experience-based preference* about which system version they would choose for future usage. User perceptions of *retrospective accuracy* with both versions of the system highly correlate with their final *experience-based system preference* in a logistic regression model (p's = [0.019, 0.0001]). Given only accuracy scores from both versions, we can predict the version choice with 69.5% accuracy in a 5-fold cross-validated test. Therefore, knowing users' *predictive accuracy* was higher for the word-level version, we would expect that this would lead to users ultimately preferring the more transparent word-level version of the system.

*After experiencing the system there is no preference for either system version.* However, this positive evaluation of word-level accuracy did not persist after actual experience with the system, when we analyzed final *experience-based preferences*. As we expected, many of those who preferred the word-level highlighting did so because it illustrated the inner workings of the system. Participant 72 said *"I know what the first* [word-level] *version is doing. I cannot tell what the second version* [document-level] *is doing. Because the second version does not give real feedback, I cannot make an informed decision when writing if I should be using it or not."* Participant 28 concurred, saying *"I think it* [the word-level version] *provides more engaging feedback and helps me better understand the reasons it gives for the amount in the meter."*

However, to our surprise, after experience with both systems overall users were evenly split in which version of the system they preferred to use in the future: Fifty percent of participants (37) said that they would prefer the document-level version if they were to use the system again. The other 50% (37) chose the word-level condition. Consistent with these preference judgments, participants also showed no overall differences in trust after using the two different system versions (t(73)=.910, p=.343). versions. While users seemed to have better impressions of the word-level condition initially, those preferences disappear after using both systems. It is important to note in this context that although the only differences between the systems lay in their transparency, users seemed to treat them as operating quite differently. As we will see later, even though participants knew both versions of the system existed at the start of the experiment, they were prone to attribute different qualities to each after experiencing them.

Overall, both the final system choice and trust showed no differences between system versions despite people being confident initially that the Word version would be more accurate. What could explain these changed perceptions after usage?

*Role of Cognitive Load:* One possible explanation for this changed perception is cognitive load. Incremental word-level feedback may demand attention and distract participants, in contrast to the document-level system version which does not present as much information. Some participants' explanations for their *experience-based preference* seem to support this. These participants (n=13) cited the distracting nature of the word-level highlighting as a motivation for preferring document level transparency. Users called word-level feedback "*annoying*" (P3) and "*obtrusive*" (P7). Participant 20 said that the document-level was "*a lot less distracting*". P57 said *"The individual highlighting of the words was distracting during writing; I wouldn't have minded it as much if I could turn it on and off."* However, these subjective reports were not borne out by our quantitative analysis of cognitive load as assessed by the TLX survey. A paired t-test comparing the overall TLX measures for both versions of the system indicated no difference in workload: t(73) = -.05, p=.95.

*Reduced transparency may lead users to overestimate system capabilities:* Another potential reason why users may prefer document-level feedback relates to user inferences about algorithmic capability. Our qualitative analysis of *experience-based preference* suggests that when users know less about the working of the system, they seem to ascribe more advanced abilities to it. Nearly a quarter (24%) of users who chose document-level transparency as their preferred version stated that they preferred it because it took their overall writing context into account, incorporating information beyond simple lexical weightings. Participant 66 said, *"I think the second* [document-level version] *one takes into account everything you are writing and makes a decision better than just by focusing on word choice."* Participant 17 concurred, saying *"I like the second* [document-level version] *as it seems to focus on the whole and not each word."* While these document-level inferences are positive, they are also inaccurate. Recall that both systems use the same underlying machine learning model which uses solely individual word features.

Contributing to this overestimation of system capabilities may be the fact that document-level feedback hides low-level errors from users. In contrast, many word-level users commented on highlighted words they felt were misclassified, leading them to downgrade their system evaluation. For example, participant 40 chose the document-level version, justifying it by saying: "*...the biggest reason is that the most negative thought I had was expressed by the word "isolated" in the text I wrote and the e-meter marked that one word as "Unimportant" I couldn't*

*get past that.*" Participant 70 said: "*Some associations don't make any sense, while others do.*" In contrast, document-level feedback did not expose these errors. If the algorithm was behaving consistently with their overall expectations, users in the document-level condition judged it very positively.

Together these observations suggest that error hiding and the absence of detailed information in document-level feedback leads some participants to form approximate but positive working heuristics about how the system operates.

**Discussion**

Our initial hypothesis was that providing detailed, incremental word-level feedback would be more helpful to users than general document-level information. However, our first study unearthed some unexpected findings, showing that user interpretations of transparency feedback are far from straightforward. Consistent with our initial expectations and the prior literature on transparency, [26,46], users anticipated a preference for word-level feedback before using the system. Users justified this preference by making arguments that such feedback would provide detailed incremental information about algorithmic decisions. But to our surprise, many participants did not retain this preference after using the system, at which point participants were evenly split between the systems in their trust and transparency preferences. As we had originally anticipated, some users continued to prefer the word-level version because of the greater transparency it provided, and they were likely to cite the increased insight that it facilitated into the algorithm's underlying operation. In contrast, others preferred document level feedback, but offered very different reasons for their preferences. Many of these users chose the document-level feedback because they seemed to find highlighting to be distracting, although our cognitive load results do not support this. Others may have preferred document-level feedback because it did not expose word-level errors, potentially leading users to overestimate the competence of the underlying algorithm with the consequence that they believed it to be more advanced than it was. For these users who preferred the document-level version, it seemed that incremental feedback was providing more information than they required [12,40].

**STUDY 2**

Ideally a system should mitigate the negative distracting elements of incremental transparency while providing improved understanding to users. However, it is difficult from our initial study to know how to operationalize transparency in a way that achieves this. In order to better understand how to convey transparency to users in effective ways we employed a semi-structured interviewing process in our second study.

We also used think-aloud interviewing methods to examine in depth how the type and timing of algorithmic transparency can inform decisions about how to design effective transparency. In particular, given that Study 1 indicated that some users felt incremental feedback provided too much information, Study 2 examines letting users view increased transparency only when they explicitly request it, after they have finished writing. Overall Study 2 gathered richer contextual qualitative data to illuminate what factors influence the interpretation and uptake of transparency information. In particular we wanted to better understand why ostensibly richer feedback was not providing anticipated benefits to some participants.

**Method**

*Users*

Twelve users were recruited from an internal participant pool at a large United States west-coast university. They received course credit for participation. Participants average age was 19.54 years (sd=1.52) and 7/12 identified as female. This study was approved by an Institutional Review Board.

*Measures*

Users completed a shortened version of the PGWBI to screen for mental health before participants began the study [30]. Users answered a similar set of survey questions to Study 1 in a think-aloud style; however, these were primarily used to prompt explanation and structure the interview. As such, we do not present the results in this paper.

*Procedure*

The participants were randomly divided into one of two conditions. Both groups were given document-level affective feedback from the E-meter (Fig 1).

Condition 1: Six participants received real-time incremental word-level feedback about the algorithm's interpretation of their affect as they typed each word.

Condition 2: The other six only obtained word-level feedback after they had finished the writing task; these users explicitly requested word-level feedback by clicking a button labeled "How was this rating calculated?".

The researcher explained the experiment and think-aloud procedure, demonstrating a think-aloud on an email client. The researcher asked participants to "Please write at least 100 words about an emotional experience that affected you in the last week." After the think-aloud writing exercise, the experimenter conducted a semi-structured interview that included an on-screen survey. After the survey, participants in the initial document-level condition 1 saw exactly the same screen with an added button labeled "How was this rating calculated" which they pressed to reveal word-level highlighting. The entire process took around 50 minutes.

*Analysis*

Interviews were recorded using both audio and screen-recording. Two interviews (one from each condition) were not audio recorded, thus only the remaining 10 are used for the analysis. For this qualitative analysis, responses were coded using thematic analysis [11] specifically targeting RQ3: What problems must be mitigated to support effective transparency?

**Results**

*More Transparency is Not Always Better:* The word-level version of the E-meter again operationalized transparency using color highlighting to show how each word contributes to the calculation of the overall mood score. However, several users took issue with this level of detail. P10 felt that only the "*big emotionally heavy words*" should be highlighted. Other users felt similarly, P4 talked about select "*trigger*" words that that "*trigger the foundation of the issue*" and were essential to understanding the text. P8 felt that there were a few important words and the rest just added noise: "*its taking into account words like 'stressful' and 'regretful' and stuff but then like everything else in between adds like an extra layer that complicates it*". While the machine learning model was limited to the 600 most predictive words of mood, that model still seemed to present too many extraneous words that users felt were unimportant. From a design perspective, it may be that users need to identify a small number of clear examples of words showing strongly positive and negative affect, in order to form a working model of system operation.

*Transparency May Violate User Expectations Even When the System is Correct:* Similar to how participants felt there were many extraneous highlighted words, participants also often focused on specific words that they judged had been misinterpreted by the system. Many users wrote about their experiences in the first person, often using the word "I". In our machine learning model, "I" has a slightly negative connotation which confused our users, because many of them thought "I" should be neutral. P8 said "*So I was gonna say that yellow words would be neutral because it has highlighted 'I'...*". Along the same lines, when analyzing the highlighting of different words, P6 said "*'I'? Mmm, I don't understand that either*". Participant 5 even started conjecturing about the actual system model saying "*'I' doesn't seem like it would have...*[participant trails off] *unless people speak in objective terms when they're talking about more positive experiences.*" Clearly these users don't feel that the word "I" has negative connotations. However, extant literature confirms that system feedback is correct as usage of first person singular pronouns are correlated with depression and negative mood [63,70]. These examples indicate a problem when system feedback reveals information that contradicts the users' expectations. Even if the system is objectively correct when displaying transparency information, it can still cause users to take issue and result in poorer perceptions of that system.

*User Heuristics Interact Negatively with Transparency:* Other problems arose because users formed working heuristics of how the system operated, which were sometimes contradicted by word-level feedback. Four of the 12 users felt that there were discrepancies between document-level and word-level feedback within the system. When Participant 8 viewed the word-level transparency they started off by saying "*I'm very confused*" and then explained how they felt the overall rating should just be calculated as a ratio of positive/negative words—"*Well I just assumed that some words would be coded as positive or negative and then it would just like do a ratio of those two.*" Participant 6 explained it similarly, the document-level rating showed a slightly negative emotion rating for their written passage but, in participant 6's words: "*So when you look at the comparison with the meter, and you look at my paragraph itself, right? There's more words that are highlighted in green.*" These users are using simple heuristics to relate the document-level and word-level transparency. They feel that the document-level rating should reflect a simple ratio of the word-level transparency. For example, if the highlighted words are primarily green, then the overall rating should be very positive. However, in our machine learning model, a single word such as "angry" could be rated negatively enough that it would cancel out multiple mildly positive words. Some users arrive at this correct model after consciously engaging with the system. For example, recall how participant 6 talked about how they felt the word-level highlighting and document-level transparency were incongruent; however, after thinking more deeply, this participant later said: "*... it's weighing certain words, right? Because obviously these two words, right? Like "upset" and like—er yeah, "upset" really polarize the meter."* This quote demonstrates that this user has moved beyond their initial heuristic that all words are weighted equally; they are now noting how one word seemed to have a larger effect in the system.

Together these observations suggest that users initially form simple working hypotheses about system operation. Users seem to engage with transparency first by operating with these simple hypotheses and only scrutinizing these when their expectations aren't met. As with other areas of reasoning, it may be that in interacting with a system users first engage in rapid, approximate, System 1 thinking and only engage in deeper, more analytic, System 2 thinking when truly prompted or confronted with clearly anomalous information [38,64].

**Discussion**

The second study again revealed the complexity involved in presenting transparency information. Confirming our first study, we again showed that providing more detailed transparency information isn't always better. While some users saw potential benefits to word level information, they argued against "complete" transparency, preferring to see only a subset of 'important' words. This observation suggests a principle for transparency presentation that limits the amount of information presented. This principle might involve explaining as much variation as possible using the smallest number of explanatory features. We might therefore aim to weigh the overall number of features we present against the information they provide. This is consistent with machine learning approaches to developing models with high dimensional feature sets, that aim to identify features with the greatest explanatory power. While we have seen technical methods that support this [53], we have not seen such transparency actually tested with users.

A second observation is that users may take issue with detailed transparency and predictions, even when underlying system models are objectively correct. This creates quite a difficult problem for system designers. If it were possible to know which features users prompt mistaken beliefs, then these features could be excluded from transparency. Unfortunately, short of testing user beliefs about all features, this may be very hard to do. We also saw that users don't actively interrogate transparency information to deeply analyze all its implications. Instead, users often look for quick heuristic routes to confirm or discredit simple working theories. We should therefore design intelligent systems in ways that allow users to develop simple working heuristics but also invite them to evolve more accurate mental models when they are motivated to do so.

**DISCUSSION AND CONCLUSIONS**

Unlike much technically oriented work that aims to develop new transparency algorithms, we explored user-centric perceptions, and reactions to, transparency. Both studies indicate that developing and deploying transparent smart systems is complex in practice, when we have users engage with real systems that interpret their personal data. In both studies, we found unexpected user reactions to our attempts to provide detailed information about algorithmic operation. In Study 1 we found that before actual usage, participants initially anticipate greater accuracy for the word-level transparency version of the E-meter but this is altered by their system experiences. After *using* both systems, they are split 50-50 in their preference, and trust data showed similar ambivalence. We identified several possible reasons for this shift. As anticipated, participants who preferred the word-level version valued the increased system understanding that transparency afforded. But two different reasons led to users preferring the document-level version: users attributed more sophisticated abilities to the less transparent system and felt distracted by increased transparency. Although many user comments mentioned how incremental word level affective feedback was highly distracting, these comments were not reflected by a reliable overall difference in cognitive load between transparency conditions as assessed by NASA TLX. However more sensitive online measures of cognitive load might yield different results.

These initial results raise the question of how we can operationalize transparency in ways that don't distract, while simultaneously allowing users to engage with the system using simple heuristics and facilitating advanced understanding. Study 2 gathered richer contextual data to illuminate exactly how to operationalize transparency to fit the goals from Study 1. To address distraction and improve clarity, we discovered that some users prefer to see only major contributing features to the document-level rating. More detailed transparency information can lead some users to falsely believe the system is operating incorrectly. Furthermore, users often evaluate transparency using simple heuristics rather than deep reflection. Together these results indicate that supporting transparency is complex and there are myriad decisions that affect the user experience when deciding how to operationalize it. We now suggest design approaches that build on these observations.

**Meeting the Competing Needs of Transparency Through Progressive Disclosure**

Studies 1 and 2 unearthed requirements that transparency must meet to be effective for users. This is a challenge because some of these requirements seem internally inconsistent. How can we allow users to develop system heuristics while at the same time facilitating detailed understanding for those users who value it? One design solution is suggested by an interaction that took place in Study 2. Recall that some users wrote their experience using the document-level version and only later saw the word-level transparency after clicking a button labeled "How was this rating generated?" After clicking the button to reveal word-level highlighting, Participant 11 had further questions. Quite naturally, the participant pointed at the button again and had this exchange with the researcher:

> P11: *Can I click this? Does this...?*
> I: *I don't think it shows any more than that.*
> P11: *Damn it.*
> I: *Yeah. If you were to click it, what would you expect to see more about?*
> P11: *I want bullet points to tell me why it works the way it does.*

Even after seeing the exhaustive word-level feature contributions to the document-level rating, this user still had more questions about the system's inner workings. This data point suggests an interaction paradigm that meets the competing needs these two studies have generated: progressive disclosure.

Progressive disclosure has a long history in UI research, dating back to the Xerox Star and early word processing systems [14,59,72]. The original concept involved hiding advanced interface controls; allowing users to make fewer initial errors and learn the system more effectively [14]. In other words, advanced information and explanation is provided on an 'as needed' basis but only when the user requests it. Progressive disclosure is also consistent with literature on explanation from the social sciences, which argues that in human-human interaction, explanations are 'occasioned', being provided only when the situation demands it [27,29,34,71]. We can apply the principles of progressive disclosure directly to transparency in intelligent systems. For example, similar to Study 2, the E-meter could show a "How was this rating calculated?" button. In this setting, the E-meter might start with only a document-level rating, which reduces distraction and avoids unnecessary complexity. Upon first press of the button, the E-meter shows a brief natural language explanation e.g. "This rating was calculated using the positive and negative weighting of the words you have written." On the next press, the E-meter shows a limited set of the high confidence and highly contributing words. This second press satisfies the users in Study 2 who only wanted to see the major factors. However, some users (like those in Study 1) may want yet more transparency. Another press of the button could reveal further features that contributed to the overall score. More presses could reveal details about how the data was collected or a textual summary of the machine learning model. In this progressive disclosure approach, explanation is presented as two-way communication with the user driving exactly when and how explanations are provided. Note too that because transparency is provided 'on demand' this removes confusions and inefficiencies arising from spurious, unwanted explanations, and adjusts explanations to the users' requirements. These design suggestions are consistent with recent work on folk theories in algorithmic systems [19,22] as well as a large body of social science theory; these theories show that people are content to operate with simple (often inaccurate) situational heuristics unless they are deeply invested in a decision or the situation is strikingly anomalous [38,64].

Progressive disclosure of transparency is not limited to predictions from text. For example, other researchers have examined transparency in the context of deep learning models predicting patient outcomes in the medical realm [43]. In this context, patients have a predicted diagnosis risk of a disease and a high number of features (e.g. previous medical diagnoses, number of doctor visits, type of care). Each feature can be visualized and ranked for its contribution to the overall predicted diagnosis risk score. In addition, users can compare outcomes between patients and conduct what-if analyses by modifying patient attributes and seeing how predicted risk changes. Our results suggest that exposing all this at once could overwhelm certain users, leading them to reject the tool. To operationalize progressive disclosure in this setting, we could present the predicted risk score by default with a short natural language explanation. Should the user request more information, we can incorporate visualizations of features that most contribute to the predicted risk score. A request for further information could show similar patients and previous outcomes that have informed the current prediction. Finally, as the user continues requesting increasing amounts of disclosure about a prediction, we know the user is invested enough to truly engage with the system and we can present an interactive what-if tool. This tool allows users to modify a patient's features and see how it changes the risk prediction, helping the user to consciously build an accurate mental model of the machine learning predictions. Again, the major contribution of progressive disclosure is avoiding overwhelming users with information, but instead slowly increasing transparency as they indicate a willingness to engage meaningfully with it. We believe this can improve the acceptance of intelligent systems in many realms.

**Impact for Future Transparency Research**

Another issue arising from our research is the role of individual differences. It was apparent that different users have varying reactions to, and expectations about intelligent systems. One possibility is that these differences arise from individual user traits, such as need for control [2] or need for cognition [13]. Future work might examine the relation between such individual traits and reactions to transparency, allowing designers to profile users and deploying personalized system versions.

Our results also have important methodological implications. One method used in prior studies is to provide potential system users with hypothetical scenarios describing system operation and eliciting reactions to those systems. These methods offer ways to collect controlled user data at scale [26,47]. However, results from Study 1 indicate the importance of direct user experience when making system evaluations; users' perceptions of the system were very different following actual usage compared with their projected reactions prior to usage. Care needs to be taken with usage of scenario-based methods.

**Limitations**

The current study examines the algorithmic domain of emotion regulation and clearly other contexts need to be explored. Furthermore, our deployment of a working algorithm meant that results were obtained for situations where our algorithm generated a moderate numbers of errors—future research should evaluate contexts where there are different levels of errors. Additionally, while users generated their own data in our system, results were not directly used to inform other aspects of the user's personal behavior so the costs of system errors were low. While this is appropriate for exploring understanding of initial algorithms with moderate error rates, future work might explore user reactions to transparency in more high stakes contexts. While we believe our implications regarding progressive disclosure are generalizable, our work derived these insights from one operationalization of transparency, namely a dynamic visualization of the algorithm. There are many other ways to depict how an algorithm operates including verbal explanations, concrete user exploration, and so forth [40,42,48,69].

**Conclusion**

Overall, our data suggest empirically motivated challenges in designing effective UX methods to support transparency for complex algorithms. Our results also unveil potential new research questions regarding user traits, system heuristics and workload. The current study indicates a promising design approach involving progressive disclosure which we intend to explore in future work. It is critical to answer these questions as we continue to deploy intelligent systems with increasing ubiquity and impact.